\documentclass[letterpaper, 10 pt, conference]{ieeeconf}  
\usepackage{amsmath}
\usepackage{graphicx}
\usepackage{breqn}

\IEEEoverridecommandlockouts                              

\title{\LARGE \bf
Interpreting Deep Neural Networks for Single-Lead ECG Arrhythmia Classification
}

\author{Sricharan Vijayarangan$^{1}$,  Balamurali Murugesan$^{1,2*}$, Vignesh R$^{1*}$,  Preejith SP$^{1}$,  \\ Jayaraj Joseph$^{1}$ and  Mohansankar Sivaprakasam$^{1,2}$  
\thanks{* Equal Contribution}
\thanks{$^{1}$ are with Healthcare Technology and Innovation Center (HTIC),
        Indian Institute of Technology (IIT-M), India
        {\tt\small sricharanv@htic.iitm.ac.in}}%
\thanks{$^{2}$ are with Department of Electrical Engineering,
        Indian Institute of Technology, Madras (IITM), India
        {}}%
}

\begin{document}
\maketitle
\thispagestyle{empty}
\pagestyle{empty}
\begin{abstract}

Cardiac arrhythmia is a prevalent and significant cause of morbidity and mortality among cardiac ailments. Early diagnosis is crucial in providing intervention for patients suffering from cardiac arrhythmia. Traditionally, diagnosis is performed by examination of the Electrocardiogram (ECG) by a cardiologist. This method of diagnosis is hampered by the lack of accessibility to expert cardiologists. For quite some time, signal processing methods had been used to automate arrhythmia diagnosis. However, these traditional methods require expert knowledge and are unable to model a wide range of arrhythmia. Recently, Deep Learning methods have provided solutions to performing arrhythmia diagnosis at scale. However, the black-box nature of these models prohibit clinical interpretation of cardiac arrhythmia.
There is a dire need to correlate the obtained model outputs to the corresponding segments of the ECG. To this end, two methods are proposed to provide interpretability to the models. The first method is a novel application of Gradient-weighted Class Activation Map (Grad-CAM) for visualizing the saliency of the CNN model. In the second approach, saliency is derived by learning the input deletion mask for the LSTM model. The visualizations are provided on a model whose competence is established by comparisons against baselines. The results of model saliency not only provide insight into the prediction capability of the model but also aligns with the medical literature for the classification of cardiac arrhythmia.

\indent \textit{Clinical relevance}— Adapts interpretability modules for deep learning networks in ECG arrhythmia classfication, allowing for better clinical interpretation.

\end{abstract}

\section{INTRODUCTION}

Cardiac arrhythmia are a group of conditions that can be characterized by irregularities in the heart rhythm. While certain types of arrhythmia may be life-threatening, other seemingly innocuous arrhythmia can increase the risk of stroke and heart failure. The most common tool used to diagnose cardiac arrhythmia is the electrocardiogram (ECG). The ECG is a simple and noninvasive procedure to assess the electrical activity of the heart. However, diagnosis of arrhythmia using a standard 12 lead resting ECG is complicated by the fact that certain arrhythmic beats can occur infrequently. The standard approach has been to perform offline processing of the ECG signals obtained from an online database using signal processing algorithms that highlight potential anomaly sections for an ECG technician or physician to review \cite{enseleit2006long}. The major drawback of this approach is the black box nature of the model which does not allow for clinical interpretation of cardiac arrhythmia.

In recent years, Deep Learning (DL) models have been used to solve a wide range of problems in vision and speech, showing a significant improvement in performance compared to feature extraction based approaches. For arrhythmia classification, Hannun \textit{et al}.
\cite{rajpurkar2017cardiologist} proposed a 34 layer CNN architecture which has been shown to be adept in classifying arrhythmia using 30 second single-lead ECG rhythms. Zihlmann \textit{et al}. \cite{zihlmann2017convolutional}  applied spectrogram transform to the ECG and provided it as an input to a convolutional recurrent neural network (CRNN). Murugesan \textit{et al}. \cite{murugesan2018ecgnet} proposed combining both CNN and LSTM in order to capture spatial and temporal features. While the black box approach of DL models might be adequate in many use cases, there is a need for interpretable models in sensitive domains such as medicine to understand model competency and potential failure cases. In the domain of imaging, significant progress has been made to provide interpretability to CNN models by understanding the saliency of models through works such as Class Activation Map \cite{zhou2016learning} and Grad-CAM \cite{selvaraju2017grad}. Additionally, a method to visualize predictions of a Long Short Term Memory network (LSTM) on an ECG  was proposed by Westhuizen \textit{et al}. \cite{van2017does} through learning an input deletion mask. However, the analysis is restricted to a single beat which fails in the case of rhythmic arrhythmia like Premature Atrial Contraction (PAC). An adaptation of these visualization techniques to time series data such as ECG would massively ameliorate our understanding of model saliency. To this end, the contributions of this paper are as follows:
\begin{itemize}
\item{We propose a novel adaptation of CNN saliency visualization to 1D ECG signals. To the best of our knowledge, our work is the first study applying saliency visualization of a CNN on an ECG signal.}
\item{We propose an extension of the LSTM visualization procedure to ECG signals, which is not restricted to a single beat.}
\item {We conduct rigorous analysis of the saliency maps and draw comparisons to traditional diagnosis as highlighted in medical literature.}
\end{itemize}       

\section{Methodology}
\subsection{Problem formulation and Architecture Description}
Given an ECG signal, the task is to visualize the network's activations on different types of cardiac arrhythmia. However, before visualizing the ECG signal, the proposed model would have to be adept at classification. To this end, we first constuct a classifier as given in Murugesan \textit{et al}. \cite{murugesan2018ecgnet}. This decision is corroborated in section \ref{class} by comparing the chosen network with other state of the art networks. This is necessary as previous comparison was restricted to only 3 classes.
\subsubsection{\textbf{Classification Network}}
The dataset $X=\{(x^{(1)},y^{(1)}),(x^{(2)},y^{(2)}),....,(x^{(m)},y^{(m)})\}$ consists of input ECG signal $x^{(i)}$ and label $y^{(i)}$, where $x^{(i)} \in R^{n}$ and $y^{(i)} \in \{0,1,...7\}$. The label $y^{(i)} = 0$ corresponds to the normal rhythm and the remaining represent different types of arrhythmia. 
The classification architecture as given in Murugesan \textit{et al}. \cite{murugesan2018ecgnet} consists of three networks, namely CNN, LSTM and Fully connected (FC). The input $x^{(i)}$ is fed into the CNN and LSTM networks which output feature vectors $z_{1}^{(i)} = F_{1}(x^{(i)};\theta_{1})$  and $z_{2}^{(i)} = F_{2}(x^{(i)};\theta_{2})$ respectively. The FC network takes the concatenated feature vector $z_{1}^{(i)}||z_{2}^{(i)}$ as input and outputs a vector $z_{3}^{(i)} = F_{3}(z_{1}^{(i)}||z_{2}^{(i)};\theta_{3})$. The vector $z_{3}^{(i)}$ is passed to a softmax function which normalizes the vector into a probability distribution $p(z_{3}^{(i)})$. Cross entropy is selected as the loss function of choice.

\begin{figure}[t]
    \centering
    \includegraphics[width=0.9\linewidth]{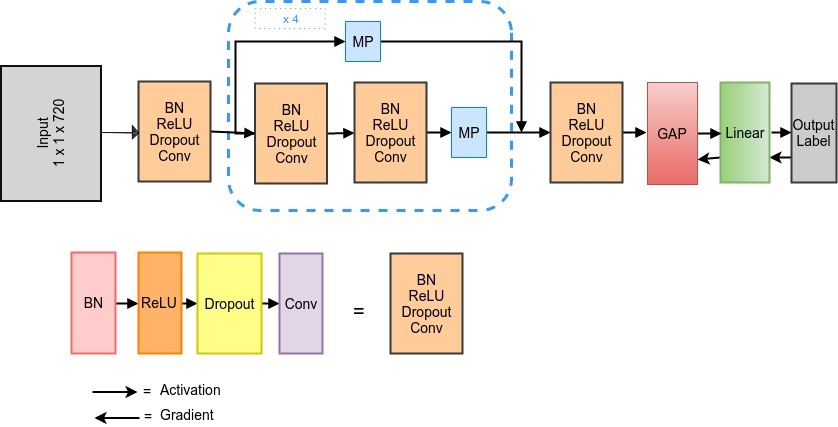}
    \caption{Schematic of CNN visualization}
    \label{CNN_Vis}
\end{figure}

The CNN used to extract features from the input ECG has two blocks, the inception block and the deep residual block. The inception block contains 4 parallel convolutional filters of varying sizes (1x15,1x17,1x19,1x21) followed by Batch Normalization (BN), ReLU, and a single convolution filter. The deep residual block consists of 15 residual units. Each residual unit contains two convolutional layers with 64$\times$k filters each of size 1$\times$16, where k is incremented by 1 for every 4\textsuperscript{th} residual unit. The LSTM is used in many-to-one fashion. For each time-step, the size of the input vector and the hidden layers are 72 and 40 respectively. The third network FC is a 3 layer deep multilayer perceptron. The networks CNN and LSTM outputs feature vectors of length 640 and 40 respectively.  These features are concatenated to a vector of length 680 and given as input to FC.  

Visualizing the CNN and LSTM networks in the above classification architecture will provide intuition on model performance by highlighting important regions of the given input ECG signals.

\subsubsection{\textbf{Visualization of CNN}}
A class activation map (CAM) for a particular label indicates the discriminative regions in an input used by the CNN to identify that label. Zhou \textit{et al}.\cite{zhou2016learning} proposed a method to obtain CAM for a specific set of networks which outputs a convolution feature map followed by Global Average Pooling (GAP) and softmax activation function. The motivation behind this method is the network's ability to preserve spatial information until the FC layers. Thus, taking the feature map before the FC layer can be used to obtain CAM. The proposed architecture is an extension of this method to ECG rhythm classification task which involves 1D convolutions. 

\begin{figure}[t]
    \centering
    \includegraphics[width=\linewidth]{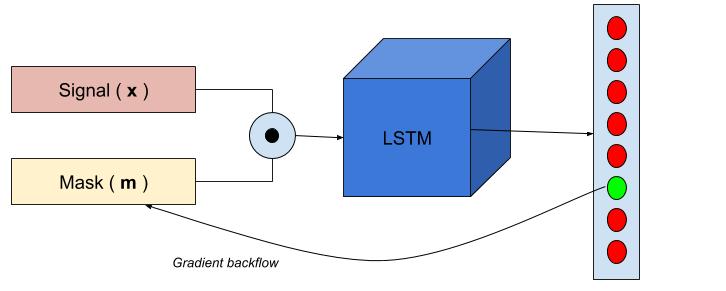}
    \caption{Schematic of LSTM visualization}
    \label{LSTM_Vis}
\end{figure}

The architecture used for CNN visualization, depicted in Fig. 1 is a miniature version of CNN and FC network in our classifier network.  The differences between the classification and visualization architectures are the reduced number of residual units, absence of inception blocks and the additional GAP unit. The visualization architecture consists of 4 residual units as compared to the 15 residual units used in the classification network. This reduction in residual units was done to alleviate the spatial information loss that occurs as a result of multiple Max Pooling operations in the residual units. The GAP unit takes these convolutional feature maps as input and outputs the spatial average of each feature map.  

For a given input signal, let $f_{k}(x)$ represent the activation of unit k in the final convolution feature map at a spatial location $x$. The spatial average output after the GAP unit is $\sum_{x}f_{k}(x)$. The output node in the FC for class $c$ is given by $\sum_{k}w_{k}^c\sum_{x}f_{k}(x)$ where $w_{k}^c$ is the weight corresponding to class $c$ for unit $k$. The CAM is defined by $M_c(x)$  where 
\begin{equation}
M_c(x) = \sum_{k}w_{k}^{c}f_{k}(x)
\end{equation}

\subsubsection{\textbf{Visualization of LSTM}}
The saliency is a set of regions in an input which upon masking, corresponds to a reduction in the predicted probability of the correct class. To obtain the saliency mask, Zeiler \textit{et al}.\cite{zeiler2014visualizing} proposed a method of occluding different portions of an input image with a grey mask. The grey mask is a rectangular box defined by location (center ($c_{x}$,$c_{y}$)) and size (width ($w$) and height ($h$)). Similarly, for signals, the mask will be a segment defined by location (center ($c_{x}$)) and size (width ($w$)). These parameters are obtained manually through repeated experiments. To mitigate this trial and error approach Westhuizen \textit{et al}.\cite{van2017does} proposed a method which learns the mask through optimization of a loss function. The method was tested in the classification of four common heartbeat classes on ECG. While this method has shown promising results, the model was evaluated only using the QRS segment of a single beat. This ignores rhythmic features among the beats, P and ST segments which play a key role in arrhythmia diagnosis. 

The LSTM visualization network ($\psi$), depicted in Figure \ref{LSTM_Vis} consists of LSTM $(F_{2})$ followed by FC network $(F_{3})$. The input to the network ($\psi$) is the element-wise product of the ECG signal ($\mathbf{x}_{1:T}$) and a mask ($\mathbf{m}_{1:T}$). The network ($\psi$) outputs softmax activations. In order to obtain the saliency mask, the loss function in Equation \ref{eq:lstm_vis_loss} is minimized. The first term in the loss function helps to localize the salient regions, the second term helps in imposing smoothness in the mask thereby removing sudden transitions and the third term helps in measuring the reduction in the network confidence score. $\lambda_{1}$ and $\lambda_{2}$ are the weights of the saliency and smoothing terms, respectively. 

\begin{multline}\label{eq:lstm_vis_loss}
J = \underset{\mathbf{m}_{1:T}}{argmin} \ \lambda_{1} \lvert \lvert \mathbf{1} - \mathbf{m}_{1:T} \rvert \rvert_{1}  +  \lambda_{2}\sum_{t=1}^{T-1}\lvert 
\mathbf{m}_{t+1} - \mathbf{m}_{t} \rvert \\ + \psi(\phi(\mathbf{x}_{1:T};\mathbf{m}_{1:T}))
\end{multline}

\begin{equation}\label{lstm_vis_model}
\phi(\mathbf{x}_{1:T};\mathbf{m}_{1:T}) = (\mathbf{1} -  \mathbf{m}_{1:T}) \odot \mathbf{x}_{1:T} + k(\mathbf{1} - \mathbf{m}_{1:T})
\end{equation}

\subsection{Dataset Description}
In this study, the ECG rhythms from the MIT-BIH Arrhythmia Database (MITDB), Long-Term Atrial Fibrillation Database (LTAFDB) and MIT-BIH Long-Term Database (LTDB) from Physionet are combined  \cite{goldberger2000current}. Lead II ECG alone was extracted from the above-mentioned datasets due to its common usage in wearable single lead ECG sensors. The beat annotations for all the datasets are provided on the R-peaks. The input windows are taken by sampling 1 second before and after the annotation, hence resulting in an input length of 2 seconds. Cardiac arrhythmia rhythms namely Normal (N), Premature Ventricular Contraction (PVC), Premature Atrial Contraction (PAC), Atrial Fibrillation (AFIB), Supraventricular Tachycardia (SVTA), Sinus Bradycardia (SBR), Left Bundle Branch Block (LBBB) and Right Bundle Branch Block (RBBB) are used. Table I summarizes the rhythm count for the given task.

\begin{table}[t]
\caption{Dataset Distribution for 8 types of heart rhythms}
\label{table_dataset1}
\begin{center}
 \begin{tabular}{||c c c c c||} 
 \hline
 Rhythm Types & MITDB & LTAFDB & LTDB & Total \\ [0.5ex] 
 \hline\hline
 N & 75013 & 10756 & 517402 & 603171\\
 \hline
 PVC & 7121 & 1318 & 5137 & 13576\\
 \hline
 PAC & 2542 & 14914 & - & 17456\\
 \hline
 AFIB & 102 & 7241 & - & 7343\\
 \hline
 SVTA & 22 & 3265 & - & 3287\\
 \hline
 SBR & - & 11323 & - & 11323\\
 \hline
 LBBB & 6580 & - & - & 6580\\
 \hline
 RBBB & 5400 & - & - & 5400\\
 \hline
 \end{tabular}
\end{center}
\end{table}

\section{Experiments}
\subsection{\textbf{Classification Model}}
Due to lack of previous benchmarks on the 8 class classification problem, three models are compared for single lead ECG arrhythmia classification namely,  Hannun \textit{et al}. \cite{rajpurkar2017cardiologist}, Zihlmann \textit{et al}. \cite{zihlmann2017convolutional} and Murugesan \textit{et al}. \cite{murugesan2018ecgnet}. The networks were implemented according to the description provided by the authors.

From Table I, it is evident that there is a severe class imbalance problem which is alleviated by undersampling the majority classes to match the number of samples in the minority class. Hence, from the available dataset, 3200 rhythms from each class are extracted and are used for training and the remaining rhythms are used to form an extensive test set. The learning rate is set to 0.005 and the momentum is set to 0.7. A mini-batch size of 16 is used and the network is trained for 30 epochs.  
\subsection{\textbf{Visualization Networks}}
For both the visualization networks, the input signals which give the correct classification with the highest confidence score are considered. Do note that the training procedure followed for classification is adapted in both CNN and LSTM visualization networks. During inference, visualization of the CNN and LSTM networks are obtained by the following methods.  

\begin{itemize}
\item For CNN visualization network, the visualization corresponds to CAM. The CAM of length 48 is obtained from the network. To overlay this map on the input signal, it is upsampled to 720.
\item For LSTM visualization network, mask $\mathbf{m}_{1:T}$ is initially set to zero. The parameters $\lambda_{1}$, $\lambda_{2}$ and learning rate are set to 1, 0.001 and 0.001 respectively. Gradient update is done using the Eq. \ref{eq:lstm_vis_loss} for 500 iterations.
\end{itemize}  

\section{Results and Discussion}
\label{class}
Table \ref{comparison} shows the quantitative results of model classification performance. It is evident that Murugesan \textit{et al}.'s model performs the best on the 8 arrhythmia classification task, which aligns with Murugesan \textit{et al}.'s observations on the 3 arrhythmia classification task. 

\begin{table}[b]
\caption{Comparisons against benchmarks}
\label{table_dataset1}
\begin{center}
 \begin{tabular}{||c c c c c||} 
 \hline
 Model & Precision & Recall & F1-Score & Accuracy \\ [0.5ex] 
 \hline\hline
 Hannun \textit{et al}. \cite{rajpurkar2017cardiologist} & 0.93      & 0.93   & 0.93     & 0.93\\
 \hline
 Zihlmann \textit {et al}. \cite{zihlmann2017convolutional}           & 0.94      & 0.94   & 0.94     & 0.94\\
 \hline
 \textbf{Murugesan \textit{et al}. \cite{murugesan2018ecgnet}} & \textbf{0.98}      & \textbf{0.97}   & \textbf{0.97}     & \textbf{0.97}     \\
 \hline
 \end{tabular}
\end{center}
\label{comparison}
\end{table}

The results of the visualization on Murugesan \textit{et al}.'s model are highlighted in Fig \ref{graph_cnn_lstm_visual}. The visualization of the network is represented as a color map where colors ranging from violet to yellow (low to high) denote the importance of a given point in a signal towards predicting the target class. The following observations can be made by studying the plots for each type of arrhythmia. 
\begin{enumerate}
\item{R peaks are observed to have a crucial role for this classification task similar to how ECG technicians infer rhythmic significance to diagnose arrhythmia \cite{hall2017guyton}. It can also be noted that for normal rhythms, LSTM visualization provides importance to P, T wave and QRS complex whereas CNN predominantly observes the QRS complex.}
\item{For premature ventricular complex rhythms, it is observed that the LSTM network provides importance to the abnormal PVC rhythm whereas the CNN network provides more importance to the proximal beats.}
\begin{figure}[t]
    \centering
    \includegraphics[width=\linewidth]{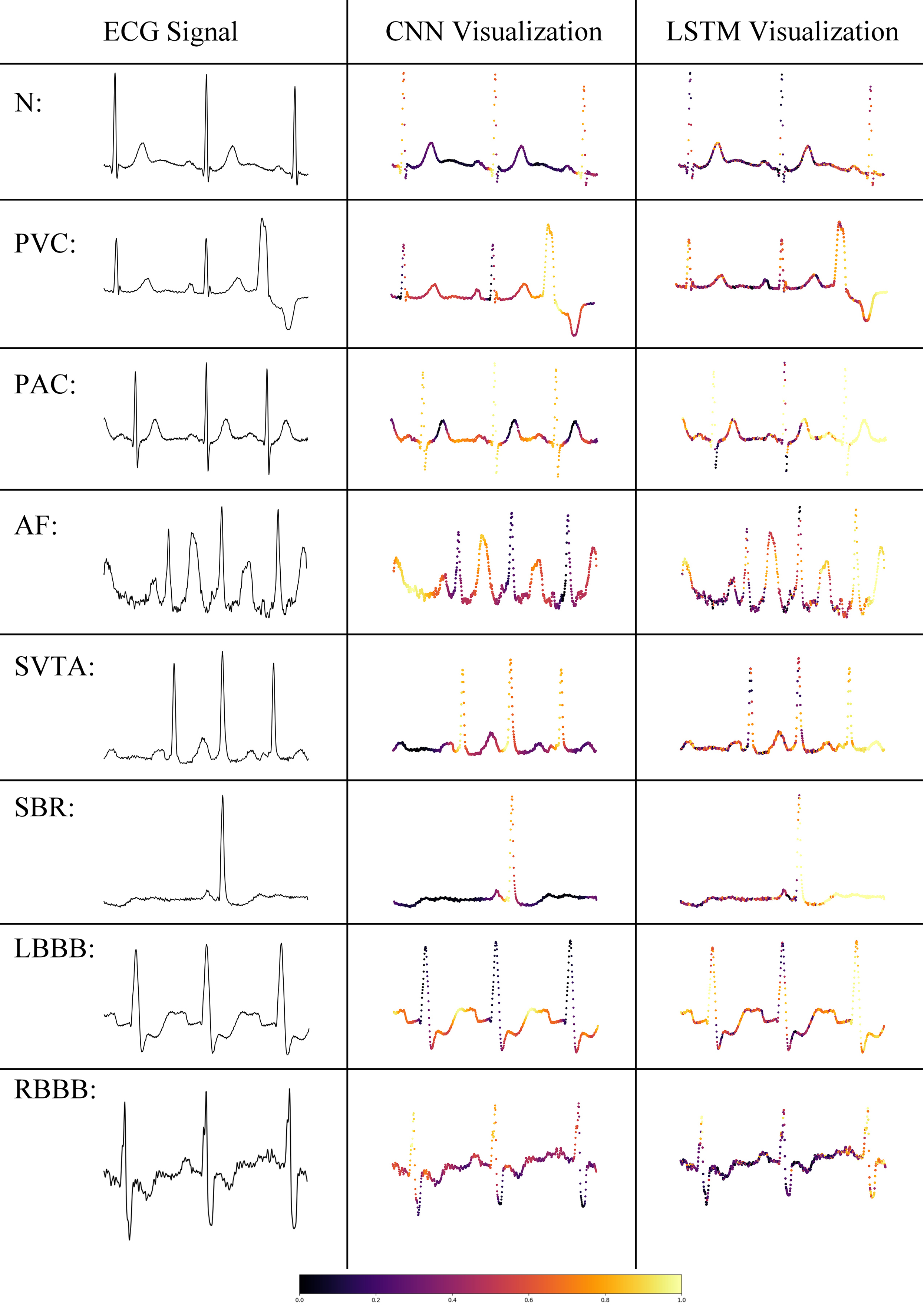}
    \caption{Visualization of CNN and LSTM for ECG signal (Scaled by importance from violet to yellow and normalized between 0 to 1). Note that the ecg samples shown are dotted and not continuous} 
    \label{graph_cnn_lstm_visual}
\end{figure}
\item{For premature atrial complex rhythms, it can be observed that the CNN network saliency is centered around the R peaks which is similar to how ECG interpretation is done for PAC using the change in inter-beat interval \cite{goldberger2017clinical}. The LSTM visualization however appears to highlight the final beat.} 

\item{By observing the visualization of CNN network for the atrial fibrillation rhythm it can be seen that the network gives importance to the R peak followed by fibrillation waves. The LSTM network too gives importance to the fibrillation waves \cite{hall2017guyton}.}
\item{While the CNN network saliency for supraventricular tachycardia rhythm appears to be focused on the QRS complex, the LSTM network focuses on the ST segment.}
\item{For sinus bradycardia it can be observed that, the CNN and LSTM network gives importance to the R peak which is similar to the standard medical practice \cite{hall2017guyton}.}
\item{The CNN and LSTM network saliency for left bundle branch block shows that the network gives more importance to the inverted ST segment and the elongated RS signal. This aligns with the medical literature for diagnosis of LBBB \cite{wagner2001marriott}.}
\item{The standard procedure for diagnosis of right bundle branch block is to identify the presence of secondary R peak and prolonged S wave.\cite{wagner2001marriott}. The CNN and LSTM network saliency gives importance to the secondary R peak during prediction.}
\end{enumerate}

\section{Conclusion}
In this work, we proposed a novel adaptation of visualization techniques of CNN and LSTM for ECG signals. The visualization was observed to line up with the clinical literature in ECG interpretation. 
Our future work will be focused on expanding the visualization to other classes and validation in a clinical setting.

\bibliographystyle{IEEEtran}
\bibliography{root}

\begin{thebibliography}{10}
\providecommand{\url}[1]{#1}
\csname url@samestyle\endcsname
\providecommand{\newblock}{\relax}
\providecommand{\bibinfo}[2]{#2}
\providecommand{\BIBentrySTDinterwordspacing}{\spaceskip=0pt\relax}
\providecommand{\BIBentryALTinterwordstretchfactor}{4}
\providecommand{\BIBentryALTinterwordspacing}{\spaceskip=\fontdimen2\font plus
\BIBentryALTinterwordstretchfactor\fontdimen3\font minus
  \fontdimen4\font\relax}
\providecommand{\BIBforeignlanguage}[2]{{%
\expandafter\ifx\csname l@#1\endcsname\relax
\typeout{** WARNING: IEEEtran.bst: No hyphenation pattern has been}%
\typeout{** loaded for the language `#1'. Using the pattern for}%
\typeout{** the default language instead.}%
\else
\language=\csname l@#1\endcsname
\fi
#2}}
\providecommand{\BIBdecl}{\relax}
\BIBdecl

\bibitem{enseleit2006long}
F.~Enseleit and F.~Duru, ``Long-term continuous external electrocardiographic
  recording: a review,'' \emph{EP Europace}, vol.~8, no.~4, pp. 255--266, 2006.

\bibitem{rajpurkar2017cardiologist}
A.~Y. Hannun, P.~Rajpurkar, M.~Haghpanahi, G.~H. Tison, C.~Bourn, M.~P.
  Turakhia, and A.~Y. Ng, ``Cardiologist-level arrhythmia detection and
  classification in ambulatory electrocardiograms using a deep neural
  network,'' \emph{Nature Medicine}, vol.~25, no.~1, pp. 65--69, 2019.

\bibitem{zihlmann2017convolutional}
M.~Zihlmann, D.~Perekrestenko, and M.~Tschannen, ``Convolutional recurrent
  neural networks for electrocardiogram classification,'' in \emph{2017
  Computing in Cardiology (CinC)}, Sept 2017, pp. 1--4.

\bibitem{murugesan2018ecgnet}
B.~Murugesan, V.~Ravichandran, K.~Ram, P.~S.P, J.~Joseph, S.~M.
  Shankaranarayana, and M.~Sivaprakasam, ``Ecgnet: Deep network for arrhythmia
  classification,'' in \emph{2018 IEEE International Symposium on Medical
  Measurements and Applications (MeMeA)}, June 2018, pp. 1--6.

\bibitem{zhou2016learning}
B.~Zhou, A.~Khosla, A.~Lapedriza, A.~Oliva, and A.~Torralba, ``Learning deep
  features for discriminative localization,'' in \emph{2016 IEEE Conference on
  Computer Vision and Pattern Recognition (CVPR)}, June 2016, pp. 2921--2929.

\bibitem{selvaraju2017grad}
R.~R. Selvaraju, M.~Cogswell, A.~Das, R.~Vedantam, D.~Parikh, and D.~Batra,
  ``Grad-cam: Visual explanations from deep networks via gradient-based
  localization,'' in \emph{2017 IEEE International Conference on Computer
  Vision (ICCV)}, Oct 2017, pp. 618--626.

\bibitem{van2017does}
J.~van~der Westhuizen and J.~Lasenby, ``Techniques for visualizing lstms
  applied to electrocardiograms,'' \emph{arXiv preprint arXiv:1705.08153}, June
  2018.

\bibitem{zeiler2014visualizing}
M.~D. Zeiler and R.~Fergus, ``Visualizing and understanding convolutional
  networks,'' in \emph{European conference on computer vision}, 2014, pp.
  818--833.

\bibitem{goldberger2000current}
A.~L. Goldberger, L.~A. Amaral, L.~Glass, J.~M. Hausdorff, P.~C. Ivanov, R.~G.
  Mark, J.~E. Mietus, G.~B. Moody, C.-K. Peng, and H.~E. Stanley, ``Current
  perspective,'' \emph{Circulation}, vol. 101, pp. e215--e220, 2000.

\bibitem{hall2017guyton}
J.~E. Hall, ``Guyton and hall: Textbook of medical physiology,'' pp. 147--149,
  2017.

\bibitem{goldberger2017clinical}
A.~L. Goldberger, Z.~D. Goldberger, and A.~Shvilkin, \emph{Clinical
  Electrocardiography: A Simplified Approach}.\hskip 1em plus 0.5em minus
  0.4em\relax Elsevier Health Sciences, 2017.

\bibitem{wagner2001marriott}
G.~S. Wagner, \emph{Marriott's practical electrocardiography}.\hskip 1em plus
  0.5em minus 0.4em\relax Lippincott Williams \& Wilkins, 2014.

\end{thebibliography}

\end{document}